\documentclass[reprint,superscriptaddress,amsmath,amssymb,aps,prd]{revtex4-2}

\usepackage{graphicx}
\usepackage{subcaption} 
\usepackage{dcolumn}
\usepackage{bm}
\usepackage[hidelinks,
            colorlinks = true,
            linkcolor = blue,
            urlcolor  = blue,
            citecolor = blue,
            anchorcolor = blue]{hyperref}
            
\usepackage[]{lineno}

\begin{document}
\title{Measurement of reactor antineutrino oscillation amplitude and frequency\\using 3800 days of complete data sample of the RENO experiment}%

\author{S. Jeon}
\affiliation{Department of Physics and Astronomy, Seoul National University, Seoul 08826, Korea}
\affiliation{Global - Learning \& Academic research institution for Master’s·PhD students, and Postdocs, Kyungpook National University, Daegu 41566, Korea}
\author{H. I. Kim}
\affiliation{Department of Physics and Astronomy, Seoul National University, Seoul 08826, Korea}
\author{J. H. Choi}
 \affiliation{Institute for High Energy Physics, Dongshin University, Naju 58245, Korea}
\author{H. I. Jang}
 \affiliation{Department of Fire Safety, Seoyeong University, Gwangju 61268, Korea}
\author{J. S. Jang}
 \affiliation{Gwangju Institute of Science and Technology, Gwangju 61005, Korea}
\author{K. K. Joo}
 \affiliation{Department of Physics, Chonnam National University, Gwangju 61186, Korea}
\author{D. E. Jung}
 \affiliation{Department of Physics, Sungkyunkwan University, Suwon 16419, Korea}
\author{J. G. Kim}
 \affiliation{Department of Physics, Sungkyunkwan University, Suwon 16419, Korea}
\author{J. H. Kim}
 \affiliation{Department of Physics, Sungkyunkwan University, Suwon 16419, Korea}
\author{J. Y. Kim}
 \affiliation{Department of Physics, Chonnam National University, Gwangju 61186, Korea}
\author{S. B. Kim}
 \affiliation{Department of Physics and Astronomy, Seoul National University, Seoul 08826, Korea}
\author{S. Y. Kim}
 \affiliation{Department of Physics, Chonnam National University, Gwangju 61186, Korea}
\author{W. Kim}
 \affiliation{Department of Physics, Kyungpook National University, Daegu 41566, Korea}
\author{E. Kwon}
 \affiliation{Department of Physics, Sungkyunkwan University, Suwon 16419, Korea}
\author{D. H. Lee}
 \affiliation{High Energy Accelerator Research Organization (KEK), Tsukuba, Ibaraki, Japan}
\author{H. G. Lee}
 \affiliation{Department of Physics, Chonnam National University, Gwangju 61186, Korea}
\author{W. J. Lee}
 \affiliation{Department of Physics and Astronomy, Seoul National University, Seoul 08826, Korea}
\author{I. T. Lim}
 \affiliation{Department of Physics, Chonnam National University, Gwangju 61186, Korea}
\author{D. H. Moon}
 \affiliation{Department of Physics, Chonnam National University, Gwangju 61186, Korea}
\author{M. Y. Pac}
 \affiliation{Institute for High Energy Physics, Dongshin University, Naju 58245, Korea}
\author{J. S. Park}
 \affiliation{Department of Physics, Kyungpook National University, Daegu 41566, Korea}
\author{R. G. Park}
 \affiliation{Department of Physics, Chonnam National University, Gwangju 61186, Korea}
\author{H. Seo}
 \affiliation{Department of Physics and Astronomy, Seoul National University, Seoul 08826, Korea}
\author{J.W. Seo}
 \affiliation{Department of Physics, Sungkyunkwan University, Suwon 16419, Korea}
\author{C. D. Shin}
 \affiliation{High Energy Accelerator Research Organization (KEK), Tsukuba, Ibaraki, Japan}
\author{B. S. Yang}
 \affiliation{Department of Physics, Chonnam National University, Gwangju 61186, Korea}
\author{J. Yoo}
 \affiliation{Department of Physics and Astronomy, Seoul National University, Seoul 08826, Korea}
\author{S. G. Yoon}
 \affiliation{Department of Physics and Astronomy, Seoul National University, Seoul 08826, Korea}
\author{I. S. Yeo}
 \affiliation{Institute for High Energy Physics, Dongshin University, Naju 58245, Korea}
\author{I. Yu}
 \affiliation{Department of Physics, Sungkyunkwan University, Suwon 16419, Korea}
\collaboration{RENO Collaboration}
\date{\today}

\begin{abstract}
{We report an updated neutrino mixing angle of $\theta_{13}$ obtained from a complete data sample of the RENO experiment. The experiment has measured the amplitude and frequency of reactor anti-electron-neutrinos ($\bar{\nu}_{e}$) oscillations at the Hanbit nuclear power plant, Younggwang, Korea, since August 2011. As of March 2023, the data acquisition was completed after a total of 3800 live days of detector operation. The observed candidates via inverse beta decay (IBD) are 1,211,995 (144,667) in the near (far) detector. Based on an observed energy-dependent reactor neutrino disappearance, neutrino oscillation parameters of $\theta_{13}$ and $\lvert\Delta m_{ee}^2\rvert$ are precisely determined as $\sin^{2}2\theta_{13}=0.0920_{-0.0042}^{+0.0044}(\text{stat.})_{-0.0041}^{+0.0041}(\text{syst.})$ and $\lvert\Delta m_{ee}^2\rvert=\left[2.57_{-0.11}^{+0.10}(\text{stat.})_{-0.05}^{+0.05}(\text{syst.})\right]\times10^{-3}~\text{eV}^{2}$. Compared to the previous RENO results published in Ref.~\cite{PhysRevLett.121.201801}, the precision is improved from 7.5\% to 6.4\% for $\sin^{2}2\theta_{13}$ and from 5.2\% to 4.5\% for $\lvert\Delta m_{ee}^2\rvert$. The statistical error of the measurement has reached our goal and is hardly improved with additional data-taking.}
\end{abstract}

\maketitle

\section{introduction}

All of neutrino mixing angles are measured and can be described by the Pontecorvo-Maki-Nakagawa-Sakata (PMNS) matrix \cite{Pontecorvo:1957qd, 10.1143/PTP.28.870}. In other words, the three-flavor neutrino oscillation paradigm is well established. The smallest neutrino mixing angle of $\theta_{13}$ was finally determined by the reactor antineutrino disappearance \cite{PhysRevLett.108.191802, PhysRevLett.108.171803, PhysRevLett.108.131801} and now determined most accurately among the three mixing angles \cite{PhysRevLett.121.201801, PhysRevLett.130.161802}. A rather large value of $\theta_{13}$ has motivated the next round of long-baseline neutrino oscillation experiments to determine the CP phase without a neutrino factory \cite{protocollaboration2018hyperkamiokandedesignreport, Abi_2020}. A precisely measured value of $\theta_{13}$ is essential for the CP phase determination as well as for testing the unitarity of the neutrino mixing matrix. \\

The RENO Collaboration presented reactor antineutrino oscillation parameters measurements based on energy and baseline dependent disappearance, using $\sim$500 and $\sim$2200 live days of data \cite{PhysRevLett.116.211801, PhysRevLett.121.201801}. The survival probability $P_{ee} \equiv P(\bar{\nu}_{e} \rightarrow \bar{\nu}_{e})$ for reactor electron antineutrino ($\bar{\nu}_{e}$) is given by Ref.~\cite{PETCOV200294},
\begin{eqnarray}
\label{eqn:survival_probability}
P_{ee} = &&1 - \sin^2 2\theta_{13} (\cos^2 \theta_{12} \sin^2 \Delta_{31} + \sin^2 \theta_{12} \sin^2 \Delta_{32}) \nonumber\\
&& - \cos^4 \theta_{13} \sin^2 2\theta_{12} \sin^2 \Delta_{21} \nonumber\\
\approx && 1 - \sin^2 2\theta_{13} \sin^2 \Delta_{ee} \nonumber - \cos^4 \theta_{13} \sin^2 2\theta_{12} \sin^2 \Delta_{21},
\end{eqnarray}
where $\Delta_{ij}$ is defined as $1.267 \Delta m_{ij}^2 L / E_{\nu}$, with $E_{\nu}$ representing the energy of the antineutrinos in MeV, $L$ is the distance from the reactor to the detector in meters, and $\Delta m_{ee}^2$ representing the effective neutrino mass squared difference in $\text{eV}^2$, defined as $\Delta m_{ee}^2 \equiv \cos^2 \theta_{12} \Delta m_{31}^2 + \sin^2 \theta_{12} \Delta m_{32}^2$ \cite{PhysRevD.72.013009}. \\

The experiment began data collection in August 2011 and completed the detector operation in March 2023. In this paper, we report more precise values of the reactor antineutrino oscillation amplitude and frequency using the entire 3800 live days of data. This measurement improves the statistical error by 12\% and the systematic error by 18\% relative to our previous result \cite{PhysRevLett.121.201801}. The reduced systematic error comes mainly from the detailed treatment of reactor-related uncertainty and partly from the more accurately measured spectrum of cosmic-induced background. \\

\section{RENO Experiment}
The RENO experiment utilizes a high flux of neutrinos from the Hanbit nuclear power plant in Younggwang, Korea. The reactor complex comprises six reactors aligned in a line with a total thermal power of 16.8 GW. Two identical near and far detectors have been operational since June 2011. The near detector is positioned 294 meters from the center of the reactors, and the far detector is 1\,383 meters away. \\

The experiment detects reactor $\bar{\nu}_{e}$ through the inverse beta decay (IBD) process, where an antineutrino interacts with a proton in the liquid scintillator (LS) to produce a positron and a neutron. The positron annihilates after losing its kinetic energy. It emits two gamma-rays, while the neutron is thermalized and captured by gadolinium (Gd) loaded in LS (Gd-LS) to emit ~8 MeV gamma-rays. The detector comprises an inner detector (ID) and an outer detector (OD). The ID, housed in a stainless steel vessel, consists of three layers: target, gamma-catcher, and buffer. The target detects IBD reactions, while the gamma catcher and buffer mitigate escaping gamma rays and external radiation. The OD serves as a veto layer by detecting entering events and shielding against external backgrounds. The ID and OD are equipped with photomultiplier tubes (PMTs) to capture scintillation light. \\

The RENO data acquisition (DAQ) system, available from technology from the Super-Kamiokande experiment \cite{sk_elec}, includes 18 front-end boards synchronized by a 60 MHz master clock. These boards utilize QTC and TDC chips to measure and digitize the time and charge of the signals, with each board handling up to 100 kHz of events. An offline software trigger system uses PMT hit counts to generate buffer, veto, or combined triggers based on specific conditions. The system efficiently processes events with a minimal dead time, recording PMT hits and generating triggers for energy and vertex reconstruction. Further technical details about the detector system, data acquisition, and detector simulation can be found in Refs.~\cite{renocollaboration2010renoexperimentneutrinooscillation, PhysRevD.98.012002}.\\

\section{Data Sample}

\begin{table}[b!]
\caption{\label{tab:trigger} Average trigger rates over 3800 days of live time in the RENO detectors. The rate of the buffer-only trigger, which is required for an IBD event, is approximately 66 (121) Hz for the near (far) detector.}
\begin{ruledtabular}
\begin{tabular}{lcc}
\textrm{Trigger type}&
\textrm{Near [Hz]}&
\textrm{Far [Hz]}\\
\colrule
Buffer-only & 66 & 121 \\
Veto-only & 320 & 36 \\
Buffer and veto & 207 & 23 \\
& & \\
Total & 593 & 180 \\
\end{tabular}
\end{ruledtabular}
\end{table}

The RENO's data-taking commenced in August 2011, and the detector operation was completed in March 2023 after 11.6 years. Data quality checks were performed to exclude runs with unreliable detector functioning or external influences, resulting in the selection of high-quality data for analysis. Consequently, a dataset of 3307.3 (3737.9) live days was recorded for the near (far) detector. The average trigger rates over the 3800-day period are summarized in Table~\ref{tab:trigger}. The rate of the buffer-only trigger necessary for IBD candidates is about 66 Hz (121 Hz) for the near (far) detector. \\

The analog signals captured by each PMT are amplified and then converted to digital signals by analog-to-digital converters (ADC). The ADC charge values are converted into pC and photoelectron (p.e.) units. A conversion factor is approximately 1.6 pC from a single photoelectron response, using a $^{137}$Cs source. The energy of an event is determined by the total collected charge ($Q_\text{tot}$), defined as the sum of the charges observed by hit PMTs within a time window of -100 to +50 ns of a trigger time. The hit PMTs are selected by requiring charges exceeding 0.3 p.e. The charge measurement in this way minimally suffers from the influence of dark hits, flashing PMT hits, and negative charges. The event vertex ($\vec{r}_\text{vtx}$) is defined as a charge-weighted average position of all active PMTs \cite{event_reco}. Reconstructed energy and vertex exhibit time-dependent and position-dependent offsets from the true energy and vertex values, primarily attributable to the geometrical characteristics of the detector and fluctuations in its response over time \cite{Kim_2023}. Several adjustments are applied to achieve the final reconstructed energy and vertex with corrected offsets, as detailed in Ref.~\cite{PhysRevD.98.012002}.\\

\section{Energy Calibration}
Accurate energy measurement is crucial for determining $\Delta m_{ee}^{2}$ and $\theta_{13}$. To calibrate the energy scale, radioactive sources with activities at or below the $\mu\text{Ci}$ level are utilized, specifically: $^{137}$Cs, $^{68}$Ge, $^{60}$Co, $^{210}$Po, $^{9}$Be, and $^{252}$Cf. These sources are contained in acrylic enclosures during data collection. Source data acquisition was performed regularly, and the observed charges were adjusted for variations in gain, charge collection efficiency, and the attenuation length of the scintillator using the peak energies associated with neutron captures. The corrected charges are averaged to represent $Q_\text{tot}$ for the peak energy of a $\gamma$-ray source. The total charge, $Q_\text{tot}$, measured in p.e., is converted into the corresponding absolute energy in MeV through a charge-to-energy conversion function derived from multiple calibration samples of sources and neutron captures. The conversion function from $Q_\text{tot}$ to the energy deposited by a positron is established from the peak energies of these $\gamma$-ray sources. Further details on the energy calibration can be found in Ref.~\cite{PhysRevD.98.012002}. \\

\section{Backgrounds}
The experiment encounters various backgrounds, including gamma rays from surrounding materials, spallation products from cosmic-induced muons, flashes from PMTs, and electronic noises. Accidental backgrounds arise from a random coincidence between a prompt-like event, mostly caused by radioactivity, and a delayed neutron-captured event and can be minimized by spatial and temporal correlation criteria. Correlated backgrounds include fast neutrons produced by cosmic muons interacting with the surrounding rocks. A fast neutron can mimic an IBD event by creating a proton recoil and a delayed neutron capture on Gd. In addition, unstable $^{9}$Li and $^{8}$He isotopes are generated inside the detector as spallation products of high-energy cosmic muons and subsequently decay with a beta emission accompanied by a neutron. An unexpected background comes from contamination by $^{252}$Cf, accidentally introduced into the liquid scintillator during a calibration procedure in 2012.

\section{Event Selections}
Event reconstruction and energy calibration can be found in Ref.~\cite{PhysRevD.98.012002}. Event selection criteria are applied to identify IBD candidate events and remove their backgrounds efficiently. Because of requiring a delayed signal from neutron capture on Gd, the fiducial volume of this analysis encompasses the entire target region, independent of an event location, and thus, the IBD detection efficiency is enhanced by 1.4\% due to spill-in IBD events from the gamma-catcher. \\

\begin{table}
\caption{\label{tab:muonveto_criteria} Timing veto criteria as a function of muon deposited energy. They are constructed by considering the deposited energy in the ID ($E_\text{ID}$) and the number of hits in the OD ($N_\text{OD}^\text{hit}$). A deposit-energy-dependent veto time interval effectively rejects muon-induced backgrounds.}
\begin{ruledtabular}
\begin{tabular}{lc|lc}
\multicolumn{2}{c|}{Near} & \multicolumn{2}{c}{Far}\\
\textrm{$E_\text{ID}$ [GeV]}&
\textrm{veto [ms]}&
\textrm{$E_\text{ID}$ [GeV]}&
\textrm{veto [ms]}\\
\colrule
$1.6<E_\text{ID}$ & 800 & $1.5<E_\text{ID}$ & 1000\\
$1.4<E_\text{ID}<1.6$ & 300 & $1.3<E_\text{ID}<1.5$ & 800\\
$1.3<E_\text{ID}<1.4$ & 200 & $1.1<E_\text{ID}<1.3$ & 500\\
$1.1<E_\text{ID}<1.3$ & 50 & $0.85<E_\text{ID}<1.1$ & 100\\
$0.07<E_\text{ID}<1.1$ & 1 & $0.07<E_\text{ID}<0.85$ & 1\\
$0.02<E_\text{ID}<0.07$ & 1 & $0.02<E_\text{ID}<0.07$ & 1\\
~~\& $N_\text{OD}^\text{hit}>50~[\text{p.e.}]$ & & ~~\& $N_\text{OD}^\text{hit}>50~[\text{p.e.}]$ & \\
\end{tabular}
\end{ruledtabular}
\end{table}

Before forming an IBD pair of prompt and delayed candidates in time, three pre-selection criteria are imposed: (i) $Q_\text{max}/Q_\text{tot}<0.07$, where $Q_\text{max}$ is the maximum charge among the ID PMTs and $Q_\text{tot}$ is the total charge summed over all PMTs, aiming at mitigating contamination from external gamma rays and flasher signals from the PMTs; (ii) $Q_\text{max}/Q_\text{tot}<0.07$ with an extended timing window of $-400\sim800~\text{ns}$ to further suppress residual flasher signals; and (iii) timing veto criteria designed to exclude events coming from the outside, which vary depending on the energy of the muons as outlined in Table~\ref{tab:muonveto_criteria}. Based on the pre-selected single events, IBD candidate pairs are obtained by applying the following conditions:
(iv) a prompt energy requirement of $0.75<E_{p}<12~\text{MeV}$;
(v) a delayed energy requirement of $6<E_{d}<12~\text{MeV}$ considering approximately 8 MeV gamma rays resulting from a neutron capture on Gd;
(vi) a time coincidence requirement of $2<\Delta t_{e^{+}-n}<100~\mu\text{s}$, accounting for an average neutron capture time of about $26~\mu\text{s}$;
(vii) a spatial coincidence requirement of $\Delta R < 2.0~\text{m}$ to minimize remaining accidental backgrounds.
In addition to the above selection criteria, the following requirements are imposed to remove backgrounds coming from fast neutrons, multiple neutrons, and contamination from $^{252}$Cf:
(viii) timing veto requirements consisting of
    (a) if any ID or OD trigger precedes the prompt candidate, 
    (b) if an ID-only trigger follows the prompt candidate, 
    (c) if ID and OD triggers follow the prompt candidate, 
    (d) if there are other subsequent pairs, 
    (e) if another prompt candidate follows the prompt candidate, 
    (f) if another prompt candidate with $E>3~\text{MeV}$ \& $Q_\text{max}/Q_\text{tot}<0.04$ follows the prompt candidate with a spatial correlation, 
    or
    (g) if another prompt candidate with $E>3~\text{MeV}$ \& $Q_\text{max}/Q_\text{tot}<0.04$ precedes the prompt candidate with a spatial correlation; 
(ix) spatial veto requirement exclusively in the far detector if the vertex of a prompt candidate is located within a cylindrical volume of 50 cm radius, centered at $x=+12.5$ cm and $y=+12.5$ cm, with $z<-1100$ cm.\\

\begin{table*}
\caption{\label{tab:selection_timeveto} Detailed timing and spatial veto requirements to eliminate the background due to the $^{252}$Cf contamination. They differ between the near and far detectors.}
\begin{ruledtabular}
\begin{tabular}{lcccc}
& \multicolumn{2}{c}{Near} & \multicolumn{2}{c}{Far} \\
\textrm{Selection criteria}&
\textrm{before Cf}&
\textrm{after Cf}&
\textrm{before Cf}&
\textrm{after Cf}\\
\colrule
(a) any ID or OD trigger preceding the prompt candidate & \multicolumn{4}{c}{300 $\mu\text{s}$} \\ 
(b) ID-only trigger following the prompt candidate & 200 $\mu\text{s}$ & 200 $\mu\text{s}$ & 200 $\mu\text{s}$ & 800 $\mu\text{s}$ \\ 
(c) ID and OD triggers following the prompt candidate & \multicolumn{4}{c}{100 $\mu\text{s}$} \\ 
(d) other subsequent pairs (adjacent IBD pairs) & 500 $\mu\text{s}$ & 500 $\mu\text{s}$ & 500 $\mu\text{s}$ & 1 s\\ 
(e) another prompt candidate following the prompt candidate & - & 1 ms & - & 1 ms\\ 
(f) another prompt candidate with $E>3$ MeV \& $Q_\text{max}/Q_\text{tot}<0.04$ & - & 10 s & - & 30 s\\ 
~~~following the prompt candidate with a spatial correlation & & $\Delta R < 400$ mm & & $\Delta R < 500$ mm\\
(g) another prompt candidate with $E>3$ MeV \& $Q_\text{max}/Q_\text{tot}<0.04$ & - & 10 s & - & 30 s\\ 
~~~preceding the prompt candidate with a spatial correlation & & $\Delta R < 400$ mm & & $\Delta R < 500$ mm\\
\end{tabular}
\end{ruledtabular}
\end{table*}

The criteria of (viii) (a), (b), (c), (d), and (e) are applied to eliminate candidate pairs caused by multiple neutrons or multiple interactions of a neutron with protons in the ID. The criteria of (viii), (f), (g), and (ix) are employed to eliminate the remaining background events due to the $^{252}$Cf contamination. Specifically, criteria of (viii), (f), and (g) are imposed to eliminate multiple neutron events arising from the decays of $^{252}$Cf, and more detailed conditions are given in Table~\ref{tab:selection_timeveto}. Criterion (ix) removes events in a region where events coming from $^{252}$Cf decays densely populate. Those eliminated events are believed to have settled at the bottom of the far detector target. \\

The events due to PMT light flashing are largely eliminated by a flasher removal criterion of $Q_\text{max} / Q_\text{tot} (-400\sim800~\text{ns}) < 0.07$. During the later stage of the experiment, the remaining event rate due to the flashing PMTs increased significantly due to PMT aging. To make additional reduction of rising flashing-PMT events, criterion (x) was developed using $Q_\text{max} / Q_\text{tot}$ and $Q_\text{ave}/Q_\text{max}$, where $Q_\text{ave}$ represents the average charge of neighboring PMTs relative to the PMT with the maximal charge. \\

By applying all of the selection criteria, a total of 1,211,995 (144,667) IBD candidates are obtained at $1.2 < E_{p} < 8.0$ MeV for 3\,800 days of the data sample. \\

\section{Detection Efficiency}
Some of the observed IBD signals from reactor $\bar{\nu}_e$ are expected from applying the selection criteria. Because of two nearly identical near and far detectors and a far-to-near ratio analysis, accurate and absolute detection efficiency is not necessary to be known for measuring the neutrino oscillation parameters. The correlated uncertainty of detection efficiency between the near and far detectors is expected to be canceled out by the far-to-near ratio analysis. The only uncorrelated uncertainty of detection efficiency needs to be considered for estimating the systematic errors of measured neutrino oscillation parameters. However, some backgrounds, coming from the $^{252}$Cf contamination, fast neutrons, and cosmic muon spallation products, are not identical between the near and far detectors. As described earlier, these background events are eliminated by the timing veto criteria, which differ between the near and far detectors. In this case, their detection efficiency uncertainties cannot be canceled out and must be estimated accurately for both detectors. \\

The detection efficiency is given by the DAQ efficiency, the Gd capture fraction, the spill-in effects \cite{PhysRevD.98.012002} and the IBD signal acceptance fraction of selection criteria (i), (ii), (iv), (v), (vi), and (vii). The uncertainty associated with the IBD cross section \cite{PhysRevD.60.053003} and the number of target protons \cite{PARK201345} is also included in that of the total detection efficiency as listed in Table~\ref{tab:det_eff}. Based on an identical performance of the near and far detectors within an uncertainty, each efficiency shown in Table~\ref{tab:det_eff} is obtained as a statistical-error weighted mean of the two detector efficiencies. The overall detection efficiency is $75.55 \pm 0.11 (\text{stat.}) \pm 0.13 (\text{syst.})$\% with an uncorrelated systematic error. \\

\begin{table*}
\caption{\label{tab:det_eff} Average detection efficiencies and their associated uncertainties for the selection criteria (i), (ii), (iv), (v), (vi), and (vii), along with the IBD cross-section, the number of target protons, DAQ efficiency, Gd capture fraction and spill-in effects. The overall detection efficiency is calculated as the statistical error-weighted mean of the near and far detection efficiencies.}
\begin{ruledtabular}
\begin{tabular}{lcccc}
\textrm{} & \textrm{Efficiency [\%]} & \textrm{Statistical Error [\%]} & \multicolumn{2}{c}{Systematic Error [\%]} \\
& & & \textrm{Uncorrelated} & \textrm{Correlated} \\
\colrule
IBD cross-section & - & - & - & 0.13 \\
Target protons & - & - & 0.03 & 0.70 \\
DAQ efficiency & 99.77 & 0.05 & 0.01 & 0.01 \\
$Q_\text{max}/Q_\text{tot}<0.07$ & 100.00 & 0.002 & 0.02 & 0.01 \\
Prompt energy & 98.77 & 0.03 & 0.01 & 0.09 \\
Delayed energy & 92.14 & 0.08 & 0.05 & 0.69 \\
Gd capture fraction & 84.95 & 0.03 & 0.1 & 0.79 \\
Temporal correlation & 96.59 & 0.04 & 0.01 & 0.45 \\
Spatial correlation & 100.00 & 0.004 & 0.02 & 0.01 \\
Spill-in & 101.40 & 0.05 & 0.04 & 0.66 \\
& & & & \\
Total detection efficiency & 75.55 & 0.11 & 0.13 & 1.46\\
\end{tabular}
\end{ruledtabular}
\end{table*}

The IBD signal losses due to selection criteria of (iii), (viii), (ix), and (x) differ between the near and far detectors and are presented in Table~\ref{tab:deadtime}. The frequencies of applying the selection criteria of (iii) and (viii) are proportional to muon and trigger rates, respectively, and thus, the selection criteria introduce different signal losses between the two detectors. Note that criterion (viii) employs distinct veto time periods because of more $^{252}$Cf contamination in the far detector. The overall signal loss associated with those selection criteria is ($40.592 \pm 0.013$)\% for the near detector and ($34.873 \pm 0.066$)\% for the far detector. \\

\begin{table*}
\caption{\label{tab:deadtime} IBD signal losses due to detector-dependent selection criteria of (iii), (viii), (ix), and (x) applied to reject background events coming from cosmic muons and $^{252}$Cf contamination. The veto requirements of the timing period, energy, and spatial correlation are listed with their respective signal losses.}
\begin{ruledtabular}
\begin{tabular}{lcc}
\textrm{Selection criteria}&
\multicolumn{2}{c}{signal loss [\%]}\\
\textrm{}&
\textrm{Near}&
\textrm{Far}\\
\colrule
timing criteria associated with muon & $26.383\pm0.003$ & $18.611\pm0.003$ \\ 
any ID or OD trigger preceding the prompt candidate & $13.192\pm0.001$ & $4.625\pm0.001$ \\ 
ID-only trigger following the prompt candidate & $1.206\pm0.001$ & $7.705\pm0.001$ \\ 
ID and OD triggers following the prompt candidate & $1.654\pm0.015$ & $0.188\pm0.001$ \\ 
other subsequent pairs (adjacent IBD pairs) & $0.000\pm0.000$ & $0.724\pm0.026$ \\ 
another prompt candidate following the prompt candidate & $1.804\pm0.012$ & $0.696\pm0.003$ \\ 
another prompt candidate with $E>3$ MeV \& $Q_\text{max}/Q_\text{tot}<0.04$ & $1.111\pm0.002$ & $2.930\pm0.052$ \\ 
~~~following the prompt candidate with a spatial correlation & & \\
another prompt candidate with $E>3$ MeV \& $Q_\text{max}/Q_\text{tot}<0.04$ & $1.121\pm0.002$ & $2.778\pm0.061$ \\ 
~~~preceding the prompt candidate with a spatial correlation & & \\
hotspot removal (only for the far detector) & - & $1.388\pm0.037$ \\
additional flasher removal & $0.349\pm0.009$ & $0.731\pm0.039$ \\
& & \\
Combined IBD signal loss & $40.592\pm0.013$ & $34.873\pm0.066$ \\
\end{tabular}
\end{ruledtabular}
\end{table*}

\section{Remaining backgrounds}
After removing backgrounds with all of the selection criteria, residual backgrounds remain in the final IBD candidate samples. The rates and spectra of remaining backgrounds need to be determined to extract those of the IBD signal obtained from background-enriched data samples. \\

\begin{figure}
    \includegraphics[width=\linewidth]{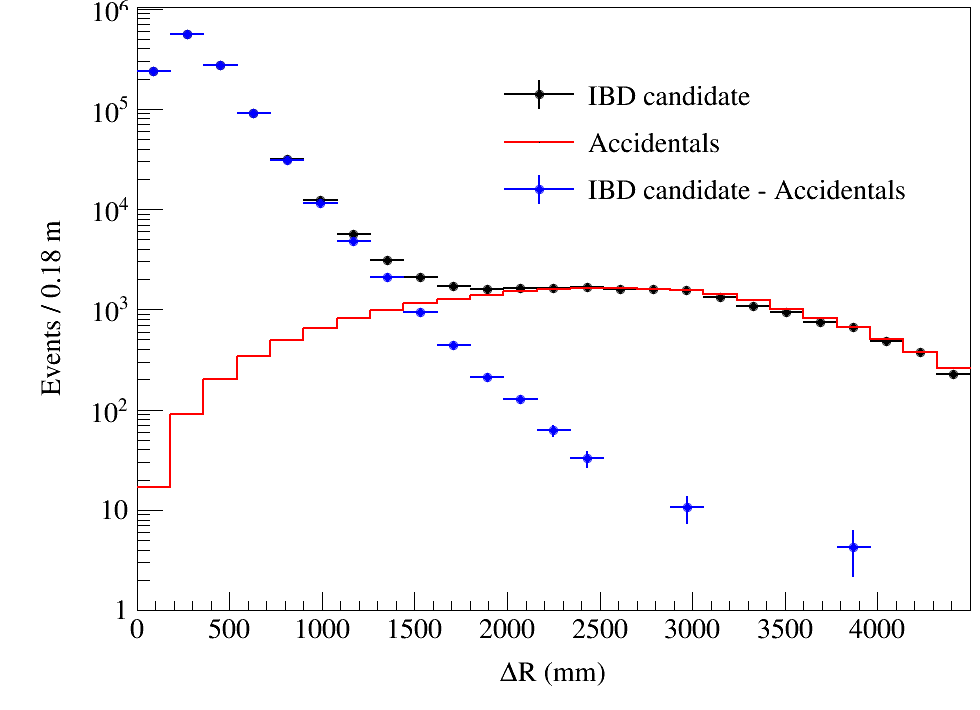}
    \caption{\label{fig:acci} Spatial correlation ($\Delta R$) distribution of IBD candidates (black dot) with no $\Delta R$ requirement. The remaining amount of accidental background in the final sample is estimated by fitting the data in the background dominant region of $\Delta R > 1.75$ m using the expected $\Delta R$ distribution (red) from the accidental background enriched sample and extrapolating the fitted $\Delta R$ distribution into the IBD signal region of $\Delta R < 2.0$ m.}
\end{figure}

\noindent{\bf{Accidental background:}}
To reduce the accidental background in the IBD candidate samples, we impose the selection criteria based on temporal and spatial correlations of $2 < \Delta t_{e^{+}-n} < 100~\mu\text{s}$ and $\Delta R < 2.0$ m between prompt and delayed candidates. An IBD candidate sample without the requirement of $\Delta R < 2.0$ m is utilized to study the remaining accidental background. An accidental background enriched sample is selected by requiring a temporal dissociation between prompt and delayed candidates, or $\Delta t_{e^{+}-n} > 1~\text{ms}$ with no $\Delta R$ requirement. It provides an expected spatial correlation, $\Delta R$, distribution of an accidental background. The $\Delta R$ distribution of an IBD candidate sample with no $\Delta R$ requirement, as shown in Fig.~\ref{fig:acci}, is fitted in the background predominant region of $\Delta R > 1.75$ m using the expected $\Delta R$ distribution of the accidental background. By extrapolating the fitted $\Delta R$ distribution into the IBD signal region of $\Delta R < 2.0$ m, the estimated remaining accidental background in the final sample is found to be $2.30 \pm 0.02$ events per day for the near detector and $0.36 \pm 0.01$ events per day for the far detector. A prompt energy spectrum of an accidental background is also obtained from the accidental background enriched sample. The fitting error of the accidental background rate determines the energy-bin-correlated uncertainty in the accidental background spectrum. The energy-bin-uncorrelated uncertainty in the accidental background spectrum comes from the statistical error of the background-enriched sample. \\

\begin{figure}
    \includegraphics[width=\linewidth]{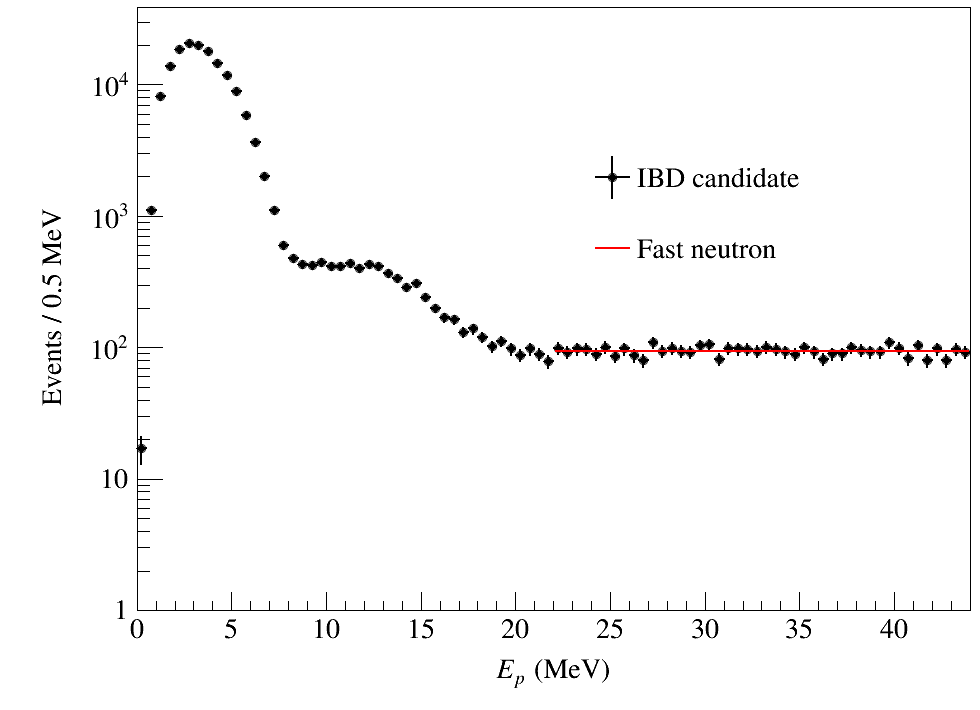}
    \caption{\label{fig:fn} Prompt energy spectrum of the IBD candidate sample for the far detector without the upper energy limit of 12 MeV. The fast neutron background rate in the IBD candidate sample is estimated by fitting the data (black dot) in the background dominant region of 22 to 46 MeV using a flat fast neutron spectrum and extrapolating the determined flat distribution (red) into the IBD signal region of 1.2 to 8.0 MeV.}
\end{figure}

\noindent{\bf{Fast neutron background:}}
The fast neutron background is estimated from an IBD candidate sample without the upper energy limit of 12 MeV on prompt energy. The remaining background rate in the final candidate sample is determined by fitting the prompt spectrum in the background dominant energy region of 22 to 46 MeV, assuming a flat spectrum of the background, as illustrated in Fig.~\ref{fig:fn}. For the sample before the $^{252}$Cf contamination, the fitting range is extended to 12 MeV. The remaining fast neutron rate is estimated to be $1.74 \pm 0.01$ (near) and $0.34 \pm 0.01$ (far) events per day by extrapolating the flat fit into the IBD signal region of 1.2 to 8.0 MeV. The background rate uncertainty is obtained from the fitting error of the flat spectrum. The spectral shape uncertainty of the fast neutron background includes a possible deviation from the flat spectrum and is added to the background rate uncertainty. \\

\begin{figure}
    \includegraphics[width=\linewidth]{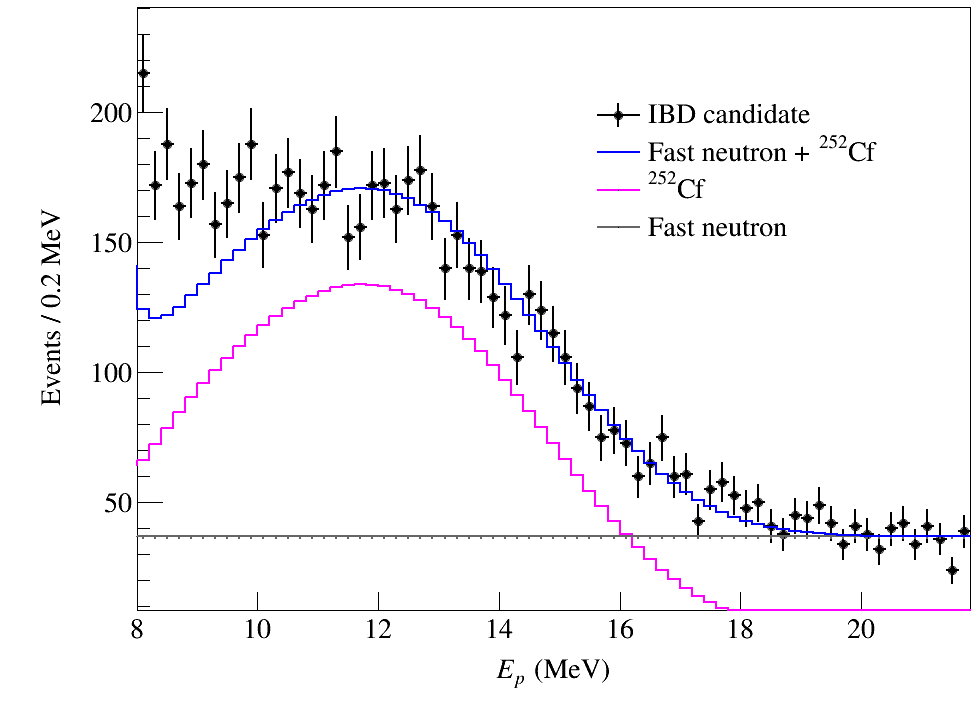}
    \caption{\label{fig:cf} Estimation of the remaining $^{252}$Cf background rate in the signal region of 1.2 to 8.0 MeV. The prompt energy spectrum of the IBD candidate sample for the far detector is fitted in the background dominant region of 12 to 22 MeV using the $^{252}$Cf contamination plus fast neutron background spectrum (blue). The remaining background rate of $^{252}$Cf contamination in the final sample is estimated by extrapolating the measured amplitude of the background dominant region into the signal region of 1.2 to 8.0 MeV.}
\end{figure}

\noindent{\bf{$^{252}$Cf contamination background:}}
As illustrated in Fig.~\ref{fig:cf}, the remaining background rate of $^{252}$Cf contamination is estimated by fitting the prompt energy distribution of the IBD candidate sample without an upper constraint on $E_{p}$, together with a $^{252}$Cf background spectrum obtained from a control sample. The background enriched sample includes the events selected by the $^{252}$Cf background removal criteria. The fitting is performed in the background dominant energy region of 12 to 22 MeV, alongside the flat spectrum of fast neutron background that are constrained within its rate uncertainty. The remaining background rate of $^{252}$Cf contamination in the final sample is estimated by extrapolating the measured amplitude of the background dominant region into the signal region of 1.2 to 8.0 MeV, as $0.07 \pm 0.01$ (near) and $0.34 \pm 0.04$ (far) events per day. The shape error of the measured $^{252}$Cf background spectrum gives the energy-bin-uncorrelated uncertainty. In contrast, the energy-bin-correlated uncertainty is assigned by the fit error of the background rate in the background dominant region. \\

\noindent{\bf{Cosmogenic $^{9}$Li/$^{8}$He background:}}
Similarly to the $^{252}$Cf background estimation, the remaining $^{9}$Li$/^{8}$He background rate can also be determined by fitting the prompt energy distribution of the IBD candidate sample in the background dominant region of 7 to 12 MeV. The fit function consists of a measured $^{9}$Li$/^{8}$He background spectrum, the Monte Carlo (MC) expected IBD spectrum, and the spectra of fast neutron and $^{252}$Cf backgrounds constrained within their rate uncertainties. The measured $^{9}$Li$/^{8}$He background spectrum is obtained using the background enriched sample of which events are produced by high-energy cosmic muons and selected by the background removal criteria. The background spectrum measurement is described in Ref.~\cite{PhysRevD.98.012002}. \\

\begin{figure}
    \includegraphics[width=\linewidth]{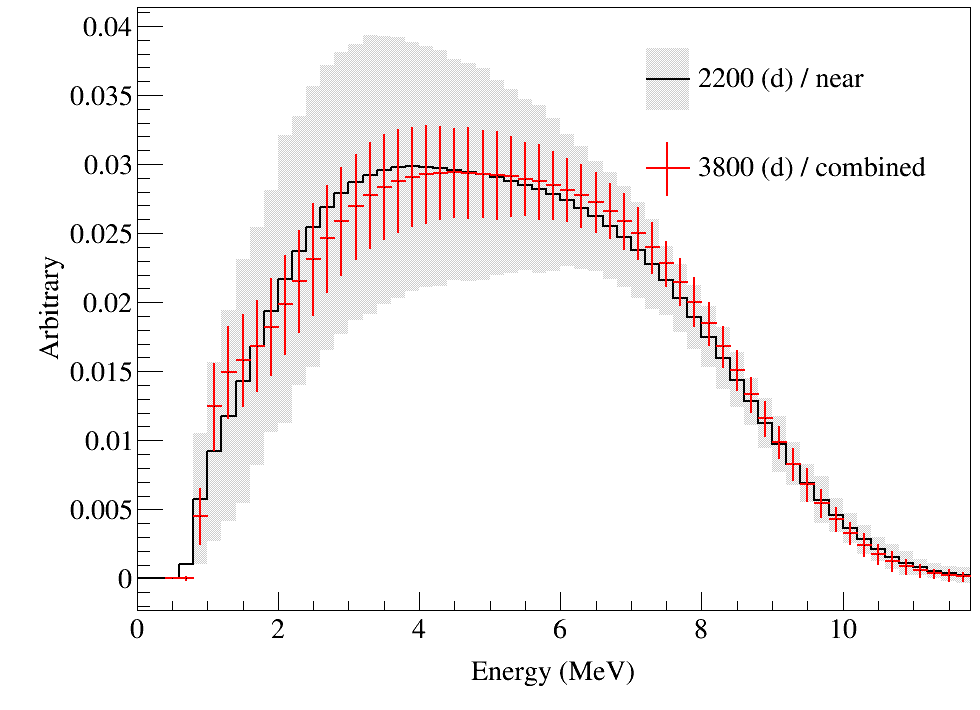}
    \caption{\label{fig:lihe_template} Measured $^{9}$Li/$^{8}$He background spectra from the background enriched samples. The updated spectrum (red solid) is obtained by combining both near and far spectra of 3800-day data. It shows significantly reduced shape errors over the previous near-detector spectrum (gray shaded) of 2200-day data \cite{PhysRevLett.121.201801}.}
\end{figure}

The uncertainty of remaining $^{9}$Li/$^{8}$He background rate constitutes a significant source of the systematic error of measured oscillation parameters. The shape error of the measured $^{9}$Li/$^{8}$He spectrum corresponds to the energy-bin-uncorrelated uncertainty, while the fitting error of the background rate leads to the energy-bin-correlated uncertainty. This analysis implements two notable improvements over the previous RENO measurements \cite{PhysRevLett.108.191802, PhysRevLett.116.211801, PhysRevD.98.012002, PhysRevLett.121.201801} in order to obtain more precise $^{9}$Li/$^{8}$He background spectrum. First, the statistics of the $^{9}$Li/$^{8}$He enriched sample is enhanced accordingly by the increased dataset from 2200 to 3800 live days. Second, the $^{9}$Li/$^{8}$He spectra separately measured at the near and far detectors are merged into a single spectrum as shown in Fig.~\ref{fig:lihe_template}, assuming their identical detector performances. The updated $^{9}$Li/$^{8}$He background spectrum exhibits a significant reduction of the shape errors in the signal region of 1.2 to 8.0 MeV, to that of 2200 live days of data, and thus results in a reduced systematic uncertainty of measured oscillation parameters. Furthermore, by using the combined $^{9}$Li/$^{8}$He spectrum for background estimation at both detectors, the correlated shape uncertainties of the two detectors are canceled out by this far-to-near ratio analysis. \\

\begin{figure}
    \includegraphics[width=\linewidth]{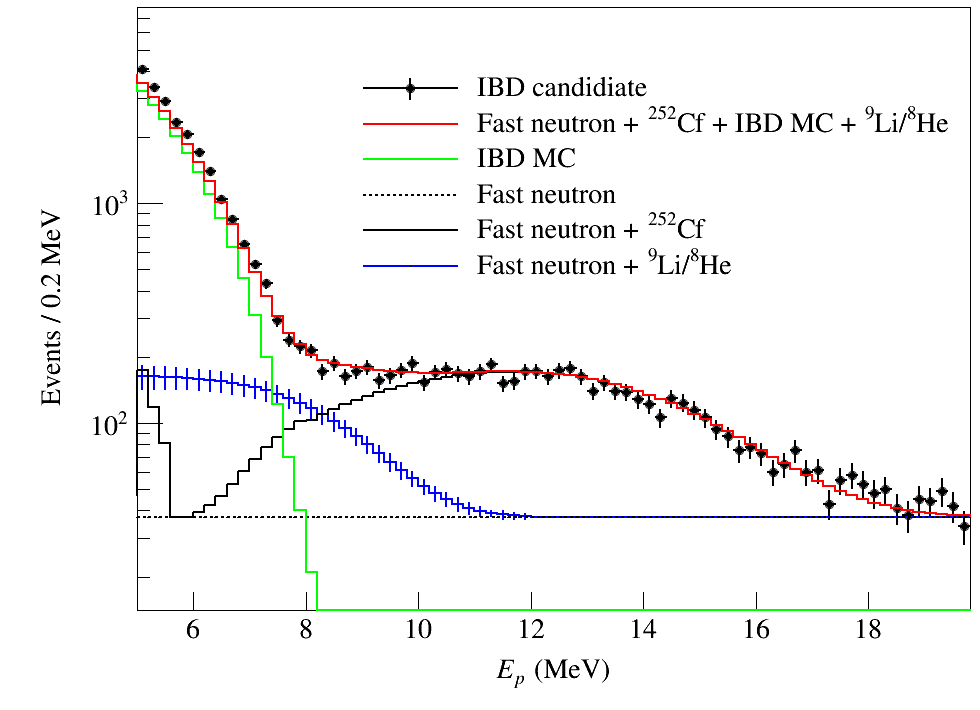}
    \caption{\label{fig:lihe} Estimation of the remaining cosmogenic $^{9}$Li/$^{8}$He background rate in the signal region using the measured rate in the background dominant region of $7 < E_{p} < 12$ MeV in the far detector. The background rate in the signal region of $1.2 < E_{p} < 8.0$ MeV is estimated by extrapolating from the measured rate in the background dominant region using the measured background spectrum.}
\end{figure}

The background rate in the IBD signal region of 1.2 to 8.0 MeV is estimated by extrapolating from the fit result in the background dominant region, using the updated $^{9}$Li/$^{8}$He background spectrum, as shown in Fig.~\ref{fig:lihe}. The remaining $^{9}$Li/$^{8}$He background rate in the final sample is estimated to be $4.97 \pm 0.17$ (near) and $1.02 \pm 0.12$ (far) events per day. The background uncertainty of the near detector is reduced to 3.4\% from 5.3\% of the previous measurement \cite{PhysRevLett.121.201801}. \\

\begin{figure}
    \includegraphics[width=\linewidth]{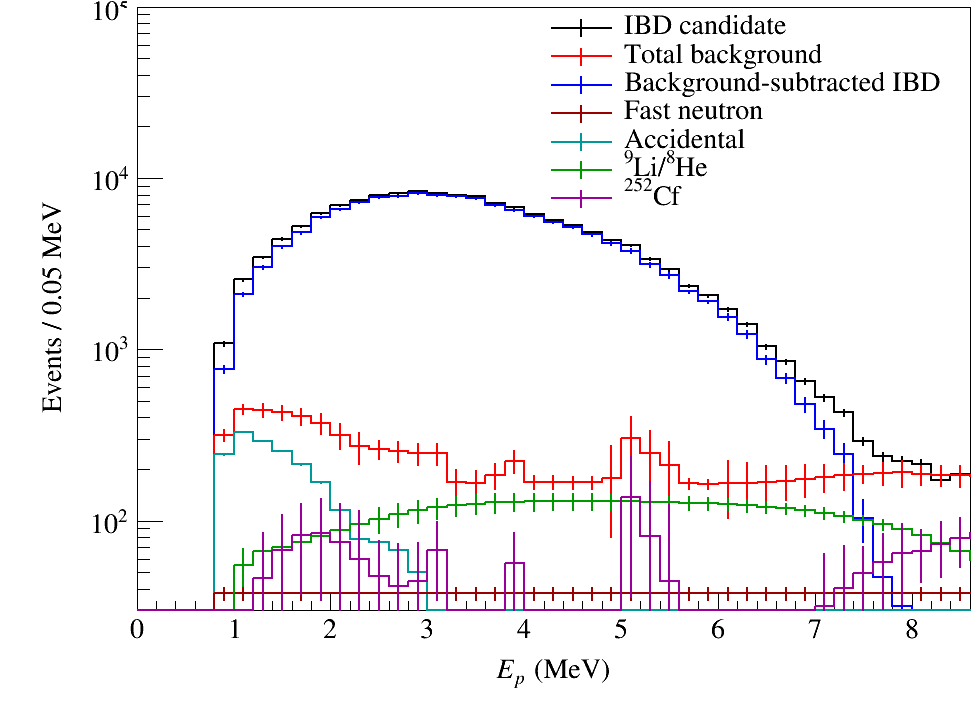}
    \caption{\label{fig:bkgsub} Observed spectrum of IBD candidates and remaining background spectra for the far detector. The total background spectrum (red) is subtracted from the final IBD candidate spectrum (black) to obtain the background subtracted IBD signal spectrum (blue).}
\end{figure}

\noindent{\bf{Background uncertainties:}}
The background-subtracted IBD spectrum is acquired by removing all the background spectra, according to their estimated rates, from the final IBD candidate spectrum, as shown in Fig.~\ref{fig:bkgsub}. Each background spectrum is measured from the background enriched sample. The total remaining background rate for $1.2 < E_{p} < 8.0$ MeV in the final IBD candidate sample is $9.08 \pm 0.18$ ($2.06 \pm 0.13$) events per day for the near (far) detector, corresponding to 2.5\% (5.3\%) of the final IBD candidates. After the background subtraction, the IBD signal rate is $357.39 \pm 0.38$ ($36.64 \pm 0.16$) events per day for the near (far) detector. Table~\ref{tab:bkgsub} summarizes the observed IBD and estimated background rates. \\

\begin{table}
\caption{\label{tab:bkgsub} Observed IBD and estimated background rates per day for $1.2 < E_{p} < 8.0$ MeV.}
\begin{ruledtabular}
\begin{tabular}{lcc}
\textrm{Detector}&
\textrm{Near}&
\textrm{Far}\\
\colrule
IBD rate after & $357.39\pm0.38$ & $36.64\pm0.16$\\
~~~background subtraction & & \\
Total background rate & $9.08\pm0.19$ & $2.06\pm0.13$\\
DAQ live time [day] & 3307.25 & 3737.85\\
\\
Accidental rate & $2.30\pm0.02$ & $0.36\pm0.01$\\
Fast neutron rate & $1.74\pm0.08$ & $0.34\pm0.02$\\
$^{252}$Cf contamination rate & $0.07\pm0.01$ & $0.34\pm0.04$\\
$^{9}$Li/$^{8}$He rate & $4.97\pm0.17$ & $1.02\pm0.12$\\
\end{tabular}
\end{ruledtabular}
\end{table}

\section{Expected and Observed IBD Rates}
The Hanbit reactors have provided almost steady thermal power during an entire cycle with a 0.5\% uncertainty per reactor provided by the Hanbit power plant and fully correlated among the reactors. The reactor $\bar{\nu}_e$ is emitted by the nuclear fission of four main isotopes of $^{235}$U, $^{238}$U, $^{239}$Pu, and $^{241}$Pu. The Hanbit power plant estimates the relative fission fractions of the four primary isotopes using the ANC reactor simulation code \cite{anc} with a quoted uncertainty of $\sim$5\% assigned to each reactor. The expected rates and spectra of reactor $\bar{\nu}_e$ for the duration of physics data taking are obtained by taking into account the varying thermal powers, fission fractions of the four fuel isotopes, energy release per fission, and fission and capture cross sections. The expected number of reactor $\bar{\nu}_e$’s in a detector is computed assuming no oscillation as Ref.~\cite{expected_ibd},
\begin{eqnarray}
\label{eqn:expected_flux}
n_{\nu}&&=\frac{N_{p}}{4\pi L^{2}}\frac{\left[\sum_{i}\alpha_{i}\bar{\sigma}_{i}\right]}{[\sum_{i}\alpha_{i}E_{i}]}P_{th}\nonumber\\
&&=\frac{N_{p}}{4\pi L^{2}}\frac{\sigma_{5}\left[1+\sum_{i}\alpha_{i}(\bar{\sigma}_{i}/\sigma_{5}-1)\right]}{E_{5}\left[1+\sum_{i}\alpha_{i}(E_{i}/E_{5}-1)\right]}P_{th},
\end{eqnarray}
where $N_{p}$ represents the number of free protons in the target, $L$ denotes the distance from the reactor to the detector, and $P_{th}$ refers to the reactor's thermal power. The index $i$ denotes each isotope of $^{235}$U, $^{238}$U, $^{239}$Pu, and $^{241}$Pu. The term $\alpha_{i}$ represents the fission fraction of the $i$th isotope, while $E_{i}$ ($E_{5}$) represents the energy released by the $i$th isotope ($^{235}$U) \cite{energy_per_fission}. Furthermore, $\bar{\sigma}_{i} = \int \sigma(E_{\nu}) \phi_{i}(E_{\nu}) dE_{\nu}$ represents the average fission cross section for the $i$th isotope, with $\sigma_{5}$ indicating the cross section for $^{235}$U. Note that $\phi_{i}(E_{\nu})$ represents the $\bar{\nu}_e$ reference energy spectrum for each isotope. \\

An expected number of reactor $\bar{\nu}_e$ events in a detector during the 3800 days are calculated by taking into account IBD cross section, data acquisition (DAQ) live time, and detection efficiency, as listed in Table~\ref{tab:expected_ibd}. \\

\begin{table}
\caption{\label{tab:expected_ibd} The expected number of reactor $\bar{\nu}_{e}$ events via IBD in the near and far detectors from each reactor, assuming no oscillation. The expected number is calculated based on reactor thermal powers, fission fractions, and the distances between each reactor and the detectors. A fractional contribution of each reactor is shown in parenthesis.}
\begin{ruledtabular}
\begin{tabular}{lcc}
\textrm{Reactor}&
\textrm{Near}&
\textrm{Far}\\
\colrule
1 & 247,328 (8.6\%) & 50,006 (15.6\%)\\
2 & 531,656 (18.6\%) & 55,875 (17.4\%)\\
3 & 926,571 (32.4\%) & 49,172 (15.3\%)\\
4 & 530,530 (18.5\%) & 39,498 (12.3\%)\\
5 & 417,591 (14.6\%) & 65,490 (20.4\%)\\
6 & 208,773 (7.3\%) & 60,639 (18.9\%)\\
\end{tabular}
\end{ruledtabular}
\end{table}

The systematic uncertainties of reactor $\bar{\nu}_{e}$ flux associated with the reactors are listed in Table~\ref{tab:reactor_syst}. The uncertainties arising from correlated sources among reactors are fully propagated to the total $\bar{\nu}_{e}$ flux, exhibiting 100\% correlation between the two detectors. Consequently, these uncertainties are canceled in the far-to-near ratio used for oscillation measurements. Thus, the only reactor-related systematic uncertainties to be considered are reactor-uncorrelated sources of baseline, thermal power, and fission fraction. The locations of two detectors and six reactors are surveyed using GPS and a total station to determine the baseline distances between the detectors and reactors with an accuracy better than 10 cm. The reactor $\bar{\nu}_{e}$ fluxes at the two detectors are calculated by assessing the flux reduction attributable to baseline distance, achieving a precision better than 0.1\%. \\

\begin{table}
\caption{\label{tab:reactor_syst} Systematic uncertainties of reactor $\bar{\nu}_{e}$ flux, uncorrelated and correlated among reactors. The uncorrelated uncertainties differ between the near and far detectors and contribute to the error of a measured neutrino oscillation parameter, while the correlated uncertainties are common for both detectors and canceled out in the far-to-near ratio analysis.}
\begin{ruledtabular}
\begin{tabular}{lcc}
\textrm{}&
\textrm{Uncorrelated [\%]}&
\textrm{Correlated [\%]}\\
\colrule
Baseline & 0.03 & -\\
Thermal power & $0.02 \sim 0.04$ & -\\
Fission fraction & 0.01 & -\\
Fission reaction & - & 1.9\\
~~~cross section & & \\
Reference & - & 0.5\\
~~~energy spectra & & \\
Energy per fission & - & 0.2\\
\colrule
Combined & $0.04 \sim 0.05$ & 2.0\\
\end{tabular}
\end{ruledtabular}
\end{table}

In the previous analyses, the uncertainty of reactor $\bar{\nu}_{e}$ flux was assigned as 0.5\% per reactor due to the uncertainty of the daily thermal power measurement and 0.7\% per reactor due to the fission fraction uncertainty. This leads to a combined reactor-related flux uncertainties of 0.9\% \cite{PhysRevD.98.012002} without taking into account the correlation from the reactor to the reactor and from cycle to cycle, although a large fraction of the uncertainties are correlated among reactors \cite{Djurcic_2009, doi:10.1142/S0217732316501200, MA2017211}. Even with the uncertainties of the thermal power and the fission fraction uncorrelated across reactors, a significant fraction of the $\bar{\nu}_{e}$ flux uncertainty is correlated between the near and far detectors. In this analysis, the correlated and uncorrelated uncertainties of the reactor $\bar{\nu}_{e}$ flux between the two detectors are estimated using a toy MC, which includes the baseline distances between reactors and detector as well as daily thermal power and fission fraction. The toy MC obtains the detector-uncorrelated uncertainties of reactor $\bar{\nu}_{e}$ flux as 0.02$\sim$0.04\% depending on a reactor due to the thermal power uncertainty, and 0.01\% due to the fission fraction uncertainty. Thus, the combined reactor-related uncertainty is significantly reduced from 0.9\% to 0.04$\sim$0.05\%. \\

\begin{figure}
    \includegraphics[width=\linewidth]{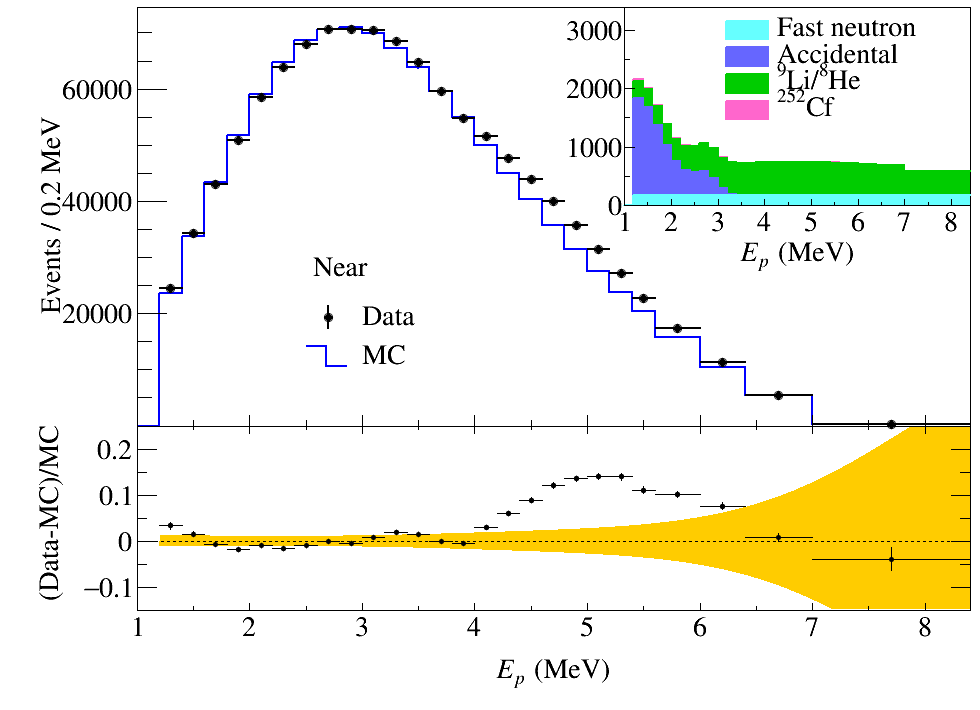}
    \includegraphics[width=\linewidth]{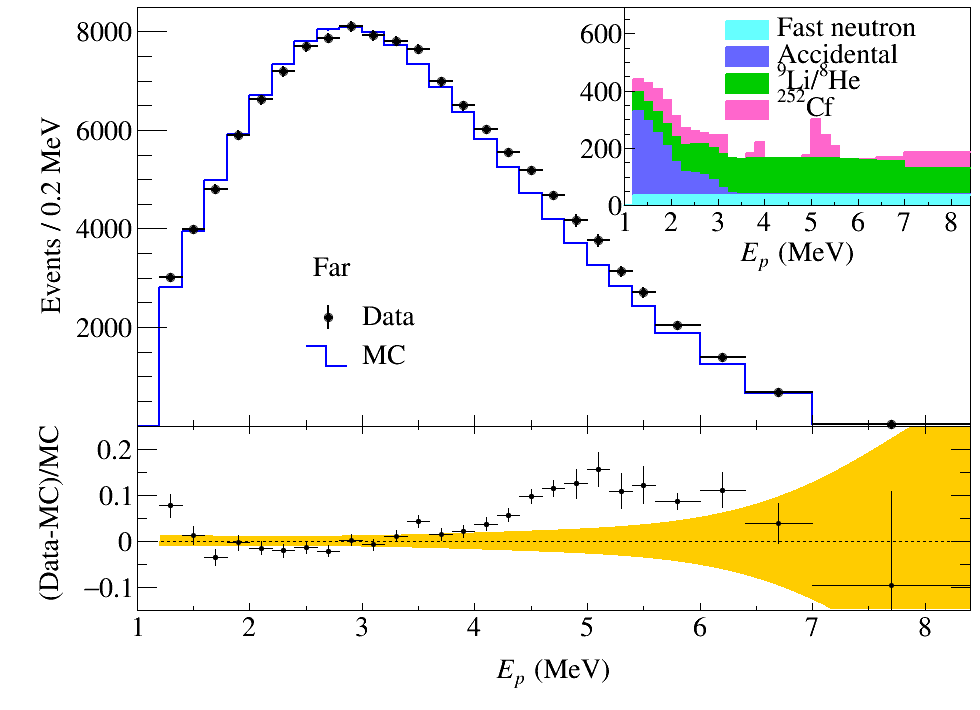}
    \caption{\label{fig:5mev} Comparison of a spectral shape between the observed and expected IBD prompt events in the near (top) and far (bottom) detectors. The observed spectra are obtained by subtracting the remaining background spectra, as shown in the insets. The expected distributions come from the best-fit oscillation results applied to the no-oscillation MC spectra. The expected spectra are normalized to the data spectra in the region excluding $3.6 < E_{p} < 6.6$ MeV. A significant discrepancy between the data and MC predictions is observed near 5 MeV. The deviation from the expectation exceeds the uncertainty of the expected spectrum (shaded band).}
\end{figure}

Fig.~\ref{fig:5mev} presents a spectral-shape comparison between the observed IBD prompt spectrum, after background subtraction, and the prediction expected from a reactor $\bar{\nu}_e$ model \cite{PhysRevC.83.054615,PhysRevC.84.024617} using the results from the far-to-near ratio measurement. The lower panel exhibits the fractional difference between the data and the predicted spectrum. A distinct discrepancy is clearly seen in the vicinity of 5 MeV across both detectors. For the spectral-shape comparison, the MC predicted spectrum is normalized to the data in the energy range excluding $3.6 < E_p < 6.6$ MeV. A clear correlation is found between the 5 MeV excess rate and the total observed IBD rate in proportion to the reactor thermal power. This suggests that the current reactor $\bar{\nu}_e$ model \cite{PhysRevC.83.054615,PhysRevC.84.024617} needs to be reevaluated and modified. \\

\section{Systematic Uncertainties}
The detector-uncorrelated systematic uncertainties of reactor $\bar{\nu}_{e}$ flux, detection efficiency, energy scale, and backgrounds are given in Table~\ref{tab:syst_uncertainty}. In the far-to-near ratio measurement, only detector-uncorrelated systematic uncertainties contribute to the uncertainties of measured oscillation parameters. The systematic uncertainty associated with the reactor $\bar{\nu}_{e}$ flux is significantly reduced by considering the correlation between both detectors, described in a prior section. Regarding detection efficiency, the uncertainties associated with the selection criteria common to both detectors are presented in Table~\ref{tab:det_eff}, and the uncertainties due to detector-dependent selection criteria are given in Table~\ref{tab:deadtime}. The uncertainty related to the energy scale is the same as estimated in the previous analyses \cite{PhysRevLett.121.201801, PhysRevD.98.012002}. The background uncertainties are presented in Table~\ref{tab:bkgsub}. \\

\begin{table}
\caption{\label{tab:syst_uncertainty} Summary of the systematic uncertainties uncorrelated between the near and far detectors.}
\begin{ruledtabular}
\begin{tabular}{lcc}
\textrm{Source} & \textrm{Uncorrelated uncertainty [\%]} \\
\colrule
Reactor & $0.04 \sim 0.05$ \\
Detection efficiency & 0.2 \\
Energy scale & 0.15 \\
Total background & 2.14 (near) \& 6.15 (far) \\
\end{tabular}
\end{ruledtabular}
\end{table} 

\section{Results}
The flux ratio measurement between the far and near detectors allows the cancellation of detector-correlated uncertainties. This paper reports the measured oscillation amplitude and frequency of neutrino survival probability based on the observed reactor $\bar{\nu}_{e}$ rates and spectra. We observe a clear energy-dependent deficit of reactor $\bar{\nu}_{e}$ in the far detector. With the spectral-shape deviation near 5 MeV from the prediction, as illustrated in Fig.~\ref{fig:5mev}, the oscillation parameters can still be determined from a fit to the far-to-near ratio of observed IBD prompt spectra. The values of $\lvert\Delta m_{ee}^2\rvert$ and $\sin^{2}2\theta_{13}$ are determined by a $\chi^{2}$ function which is constructed as the spectral ratios of observed and expected reactor $\bar{\nu}_{e}$ rates, together with pull parameter terms accounting for systematic uncertainties, and minimized by varying the oscillation and pull parameters Ref.~\cite{dc_chi2}. The $\chi^{2}$ function is written as
\begin{eqnarray}
\chi^{2}=&&\sum_{i=1}^{N_{E}} \frac{\left(O_{i}^{F/N}-T_{i}^{F/N}\right)^{2}}{U_{i}^{F/N}} + \sum_{d=N,F}\left(\frac{b^{d}}{\sigma_\text{bkg}^{d}}\right)^{2}\nonumber\\
&&+\sum_{r=1}^{6}\left(\frac{f_{r}}{\sigma_\text{flux}^{r}}\right)^2+\left(\frac{\epsilon}{\sigma_\text{eff}}\right)^2+\left(\frac{\eta}{\sigma_\text{scale}}\right)^2,
\end{eqnarray}
where $O_{i}^{F/N}$ represents the observed far-to-near ratio of IBD candidates in the $i$th $E_{p}$ bin after background subtraction, $T_{i}^{F/N} = T_{i}^{F/N}(b^{d},f_{r},\epsilon,\eta;\theta_{13},\lvert\Delta m_{ee}^2\rvert)$ denotes the expected far-to-near ratio of IBD events, and $U_{i}^{F/N}$ denotes the statistical uncertainty of $O_{i}^{F/N}$. \\

The expected far-to-near ratio $T_{i}^{F/N}$ is obtained from the reactor $\bar{\nu}_{e}$ model, the IBD cross-section, and the detection efficiency, including the signal loss due to the timing veto criteria. It also incorporates the $\bar{\nu}_{e}$ survival probability and detector responses. The systematic uncertainties are taken into account pull parameters ($b^{d}$, $f_{r}$, $\epsilon$, and $\eta$) with their uncertainties ($\sigma_\text{bkg}^{d}$, $\sigma_\text{flux}^{r}$, $\sigma_\text{eff}$, and $\sigma_\text{scale}$) listed in Table~\ref{tab:syst_uncertainty}. These pull parameters facilitate deviations from the expected far-to-near ratio of IBD events within their respective systematic uncertainties. In other words, the systematic uncertainties are propagated to determine oscillation parameters during the minimization process. The pull parameters $b^{d}$ and $\eta$ introduce changes in the expected spectra, accounting for the impact of energy-dependent systematic uncertainties. For the spectral deviations, energy-correlated and uncorrelated uncertainties are considered separately. The $\chi^{2}$ function is formed as a sum of two periods, before ($\sim$400 live days) and after ($\sim$3400 live days) the $^{252}$Cf contamination. \\

The best-fit values for the oscillation parameters are obtained as $\sin^{2}2\theta_{13}=0.0920_{-0.0042}^{+0.0044}(\text{stat.})_{-0.0041}^{+0.0041}(\text{syst.})$ and $\lvert\Delta m_{ee}^2\rvert=\left[2.57_{-0.11}^{+0.10}(\text{stat.})_{-0.05}^{+0.05}(\text{syst.})\right]\times10^{-3}~\text{eV}^{2}$ with $\chi^{2}/\text{NDF}=64.0/66$, where NDF denotes the number of degrees of freedom. Table~\ref{tab:syst_decomposition} summarizes contributions of systematic uncertainties to the systematic errors of the measured oscillation parameters. In the previous measurement from the 2200 live day data \cite{PhysRevLett.121.201801}, the uncertainties of reactor $\bar{\nu}_{e}$ flux, detection efficiency, and background are primary sources of the systematic error in determining $\sin^{2}2\theta_{13}$. The reactor $\bar{\nu}_{e}$ flux uncertainty is reduced in this analysis as described earlier, and thus its contribution to the systematic error of $\sin^{2}2\theta_{13}$ is $\pm$0.0013, significantly less than $\pm$0.0032 of the previous measurement. In this result, the energy scale uncertainty is the most dominant contribution to the systematic error of $\lvert\Delta m_{ee}^2\rvert$, sensitive to distortion in the $\bar{\nu}_{e}$ spectrum. On the other hand, the detection efficiency uncertainty contributes the largest to the systematic error of $\sin^{2}2\theta_{13}$. \\

\begin{table}
\caption{\label{tab:syst_decomposition} Contribution to the systematic errors of the measured oscillation parameters. The largest systematic error of $\lvert\Delta m_{ee}^2\rvert$ comes from the energy scale uncertainty, while the detection efficiency uncertainty is the dominant source of the systematic error of $\sin^{2}2\theta_{13}$.}
\begin{ruledtabular}
\begin{tabular}{lcc}
\textrm{}&
\textrm{$\delta \lvert\Delta m_{ee}^2\rvert \left(\times10^{-3}~\text{eV}^{2}\right)$}&
\textrm{$\delta \left(\sin^{2}2\theta_{13}\right)$}\\
\colrule
Reactor & - & +0.0012, -0.0013\\
Detection efficiency & - & +0.0031, -0.0032\\
Energy scale & +0.051, -0.051 & +0.0016, -0.0016\\
Backgrounds & +0.021, -0.021 & +0.0019, -0.0020\\
& & \\
Total & +0.055, -0.055 & +0.0042, -0.0043\\
\end{tabular}
\end{ruledtabular}
\end{table}

\begin{figure}
    \includegraphics[width=\linewidth]{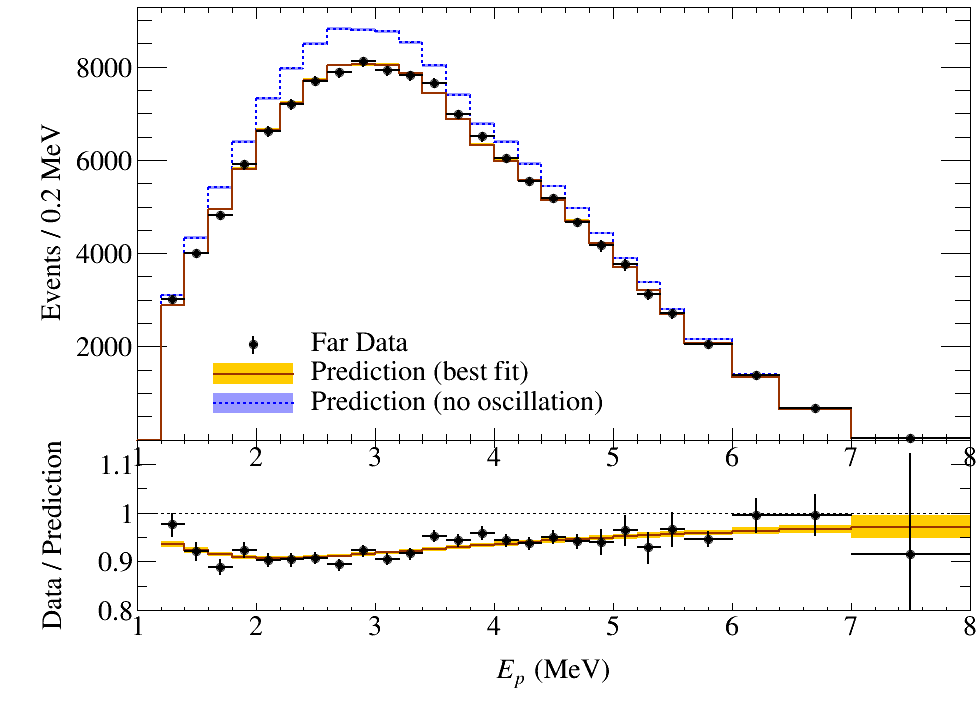}
    \caption{\label{fig:fartoprediction} Top: Comparison between the observed background-subtracted IBD prompt spectrum in the far detector (dots) and the expected spectra with (yellow shaded histogram) and without (blue shaded histogram) oscillation at the far detector. The expected spectra are obtained from the measured IBD prompt spectrum in the near detector. The bands represent uncertainties. Bottom: Comparison between the ratio of IBD events measured in the far detector to the no-oscillation prediction (dots) and the ratio from the MC simulation with best-fit results incorporated (shaded band). The errors are statistical uncertainties only, although both statistical and systematic uncertainties are considered in the $\chi^{2}$ fitting.}
\end{figure}

\begin{figure}
    \includegraphics[width=\linewidth]{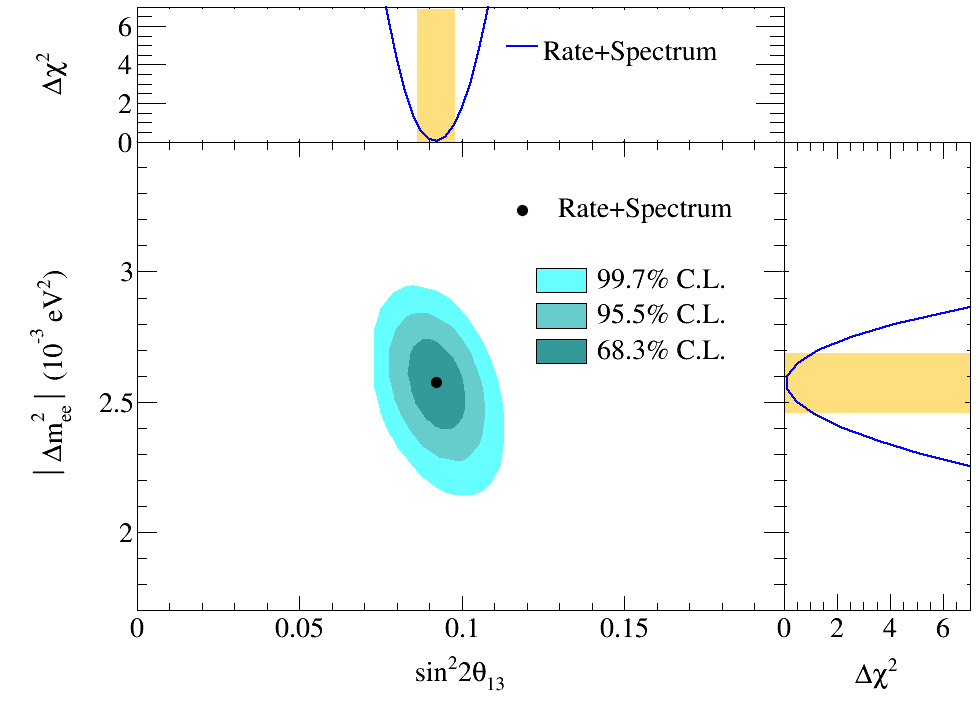}
    \caption{\label{fig:contour} Allowed regions of 68.3\%, 95.5\%, and 99.7\% confidence levels in the $\lvert\Delta m_{ee}^2\rvert$ vs $\sin^{2}2\theta_{13}$ plane. The black dot represents the best-fit values. The $\Delta \chi^{2}$ distributions for $\lvert\Delta m_{ee}^2\rvert$ (right) and $\sin^{2}2\theta_{13}$ (top) are shown with 1$\sigma$ bands (yellow shaded).}
\end{figure}

Fig.~\ref{fig:fartoprediction} compares the observed IBD prompt spectrum after the background subtraction and the expected spectra at the far detector. The expected spectrum with no oscillation is obtained by weighting the spectrum measured in the near detector with the no-oscillation assumption to incorporate the oscillation effect between the near and far detectors and the 5 MeV excess. The expected spectrum with the best-fit oscillation parameters is obtained by applying the measured oscillation parameters to the expected spectrum with no oscillation at the far detector. The observed spectrum at the far detector shows a clear energy-dependent disappearance of reactor $\bar{\nu}_{e}$ events, consistent with neutrino oscillation. Fig.~\ref{fig:contour} shows the allowed regions of 68.3\%, 95.5\%, and 99.7\% confidence levels in the neutrino oscillation parameters of $\lvert\Delta m_{ee}^2\rvert$ and $\sin^{2}2\theta_{13}$. \\

\begin{figure}
    \includegraphics[width=\linewidth]{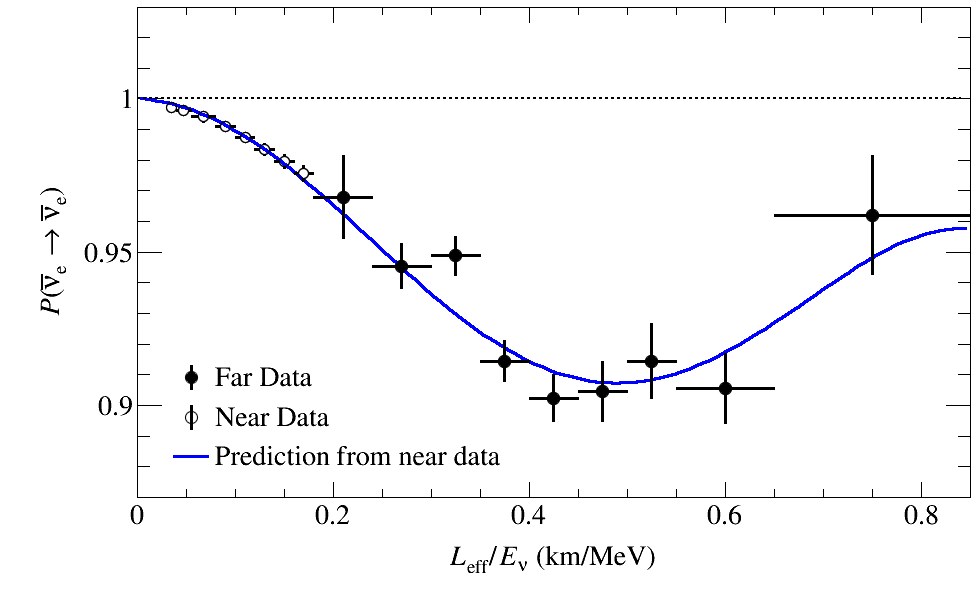}
    \caption{\label{fig:lovere} Measured reactor $\bar{\nu}_{e}$ survival probability in the far detector as a function of $L_\text{eff} / E_{\nu}$. The blue curve represents the predicted survival probability, calculated from the observed probability in the near detector, using the best-fit values for $\lvert\Delta m_{ee}^2\rvert$ and $\sin^{2}2\theta_{13}$. The $L_\text{eff} / E_{\nu}$ value for each point is determined as the average of the counts within each bin.}
\end{figure}

The survival probability of reactor $\bar{\nu}_{e}$ depends on a baseline $L$ and neutrino energy $E_{\nu}$, as given in Eq.~\eqref{eqn:survival_probability}. With multiple reactors serving as neutrino sources, an effective baseline $L_\text{eff}$ is calculated as the average reactor-detector distance weighted by the IBD event rate in a detector and expected from a reactor. Note that $L_\text{eff}$ is time dependent because it is weighted by the IBD rate. The neutrino energy $E_{\nu}$ is converted from the IBD prompt energy. A daily ratio $L_\text{eff} / E_{\nu}$ is obtained from the background-subtracted IBD spectrum combined with the daily $L_\text{eff}$. The overall distribution of $L_\text{eff} / E_{\nu}$ is obtained by collecting the daily ratios that are weighted by a daily IBD rate. The measured survival probability is determined by the ratio of the observed IBD rate to the expected rate with no oscillation for each bin of $L_\text{eff} / E_{\nu}$. Fig.~\ref{fig:lovere} presents the measured survival probability of reactor $\bar{\nu}_{e}$ at the far detector as a function of $L_\text{eff} / E_{\nu}$. A predicted survival probability is calculated from the observed distribution in the near detector and the best-fit values of oscillation parameters. Due to the observed excess near 5 MeV, the expected $L_\text{eff} / E_{\nu}$ is taken from the measured spectrum in the near detector rather than the IBD Monte Carlo spectrum. A clear $L_\text{eff} / E_{\nu}$-dependent disappearance of reactor $\bar{\nu}_{e}$ events is observed and indicates a periodic nature of neutrino oscillation. \\

\begin{table*}[t!]
\caption{\label{tab:reno_comparison} Comparison of $\sin^{2}2\theta_{13}$ and $\lvert\Delta m_{ee}^2\rvert$ values and precisions determined by this new measurement and the previous RENO measurements. A continuous improvement in precision is seen with more data and reduced systematic uncertainties. This comparison highlights the progress made by the RENO experiment in precisely determining $\theta_{13}$ and $\lvert\Delta m_{ee}^2\rvert$.}
\begin{ruledtabular}
\begin{tabular}{lcccc}
Live days & 220 \cite{PhysRevLett.108.191802} & 500 \cite{PhysRevLett.116.211801} & 2200 \cite{PhysRevLett.121.201801} & 3800 (this result)\\
$\sin^{2}2\theta_{13}$ & $0.113 \pm 0.023$ & $0.082 \pm 0.011$ & $0.0896 \pm 0.0067$ & $0.0920 \pm 0.0059$\\
~~~Precision & 20.4\% & 13.4\% & 7.5\% & 6.4\%\\
$\lvert\Delta m_{ee}^2\rvert \left(\times10^{-3}~\text{eV}^{2}\right)$ & - & $2.62 \pm 0.26$ & $2.68 \pm 0.14$ & $2.57 \pm 0.12$\\
~~~Precision & - & 9.9\% & 5.2\% & 4.5\%\\
\end{tabular}
\end{ruledtabular}
\end{table*}

\begin{figure}
    \includegraphics[width=\linewidth]{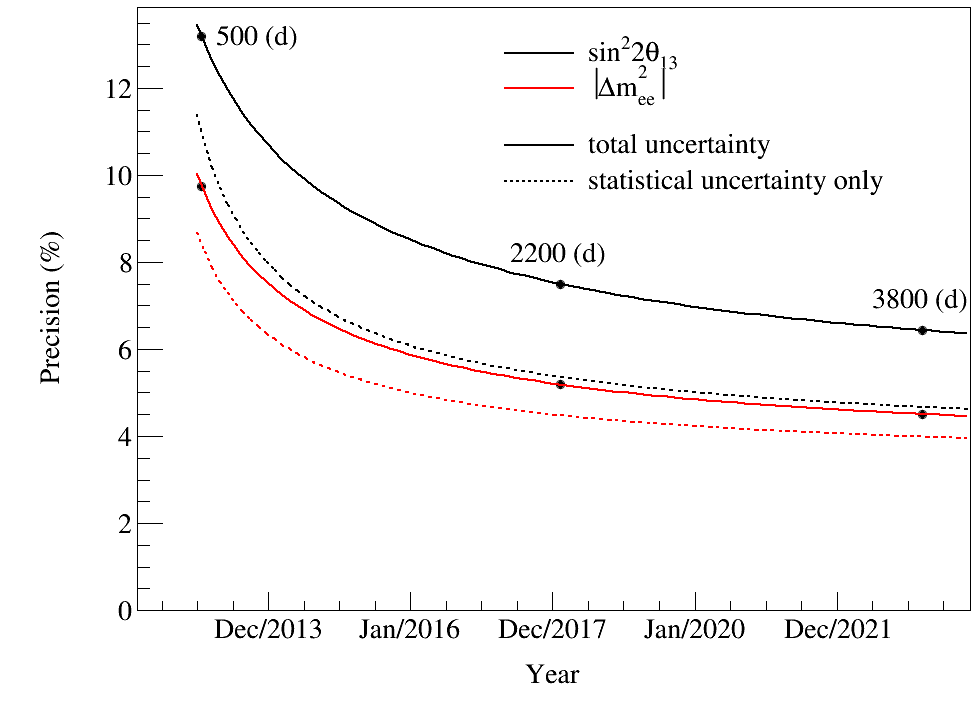}
    \caption{\label{fig:reno_comparison} Precision improvement of the measured $\sin^{2}2\theta_{13}$ (black) and $\lvert\Delta m_{ee}^2\rvert$ (red) as a function of year. The solid lines represent the total uncertainty, while the dotted lines show the statistical uncertainty only. Data points of 500, 2200, and 3800 live days indicate the precisions achieved at different stages of the RENO experiment, with a non-negligible reduction of uncertainties as more data are collected. The precision curves exhibit the experiment reaches the saturation region.}
\end{figure}

Table~\ref{tab:reno_comparison} compares the measured oscillation parameters from previous RENO measurements and this new result. A gradual improvement of precision is seen according to the increasing live days, as illustrated in Fig.~\ref{fig:reno_comparison}. This precision improvement of $\sin^{2}2\theta_{13}$ and $\lvert\Delta m_{ee}^2\rvert$ over time justifies not only a long-standing operation of the RENO experiment but also their efforts on reducing systematic uncertainties. In comparison with the previous measurement \cite{PhysRevLett.121.201801}, due to the increase of IBD statistics by approximately 40\% and an effort to reduce systematic uncertainties related to the reactor $\bar{\nu}_{e}$ flux and the $^{9}$Li/$^{8}$He background spectrum, the precision of this new measurement is improved from 7.5\% to 6.5\% for $\sin^{2}2\theta_{13}$ and from 5.2\% to 4.5\% for $\lvert\Delta m_{ee}^2\rvert$.

\section{Summary}
The RENO experiment has precisely measured the amplitude and frequency of reactor $\bar{\nu}_{e}$ oscillations at the Hanbit nuclear power plant since August 2011. As of March 2023, the RENO DAQ system has been decommissioned, resulting in a dataset of roughly 3800 live days. With this complete dataset, 1\,211\,995 (144\,667) IBD candidates are identified over 3307.25 days (3737.85 days) in the near (far) detector. Using the far-to-near ratio analysis and n-Gd captures, the reactor neutrino oscillation parameters are determined as $\sin^{2}2\theta_{13}=0.0920_{-0.0042}^{+0.0044}(\text{stat.})_{-0.0041}^{+0.0041}(\text{syst.})$ and $\lvert\Delta m_{ee}^2\rvert=\left[2.57_{-0.11}^{+0.10}(\text{stat.})_{-0.05}^{+0.05}(\text{syst.})\right]\times10^{-3}~\text{eV}^{2}$. Compared to RENO’s previous results based on $\sim$2200 live days \cite{PhysRevLett.121.201801}, the precision is improved from 7.5\% to 6.4\% for the measurement of $\sin^{2}2\theta_{13}$ and from 5.2\% to 4.5\% for $\lvert\Delta m_{ee}^2\rvert$. With $\sim$3800 live days of the RENO’s entire dataset, additional results of other physics topics are anticipated. \\

\begin{acknowledgments}
The RENO experiment is supported by the National Research Foundation of Korea (NRF) Grants No. 2009-0083526, No. 2019R1A2C3004955, No. 2021R1A2C1013661, No. 2022R1A5A1030700, and No. 2022R1A3B1078756 funded by the Korean Ministry of Science and ICT. Some of us have been supported by a fund from the BK21 of NRF. We gratefully acknowledge the cooperation of the Hanbit Nuclear Power Site and the Korea Hydro \& Nuclear Power Co., Ltd. (KHNP). We thank KISTI for providing computing and network resources through GSDC, and all the technical and administrative people who helped make this experiment possible.
\end{acknowledgments}


\nocite{*}
\bibliography{apssamp}
\end{document}